# Whose Side are Ethics Codes On?

## Power, Responsibility and the Social Good

Anne L. Washington[†]
Applied Statistics, Social Science, and Humanities
New York University
New York City, NY, USA
washingtona@acm.org

Rachel Kuo
Media, Culture, and Communication
New York University
New York City, NY, USA
rachel.kuo@nyu.edu

## ABSTRACT

The moral authority of ethics codes stems from an assumption that they serve a unified society, yet this ignores the political aspects of any shared resource. The sociologist Howard S. Becker challenged researchers to clarify their power and responsibility in the classic essay: Whose Side Are We On. Building on Becker's hierarchy of credibility, we report on a critical discourse analysis of data ethics codes and emerging conceptualizations of beneficence, or the "social good", of data technology. The analysis revealed that ethics codes from corporations and professional associations conflated consumers with society and were largely silent on agency. Interviews with community organizers about social change in the digital era supplement the analysis, surfacing the limits of technical solutions to concerns of marginalized communities. Given evidence that highlights the gulf between the documents and lived experiences, we argue that ethics codes that elevate consumers may simultaneously subordinate the needs of vulnerable populations. Understanding contested digital resources is central to the emerging field of public interest technology. We introduce the concept of digital differential vulnerability to explain disproportionate exposures to harm within data technology and suggest recommendations for future ethics codes..

## CCS CONCEPTS

• codes of ethics • computing profession • machine learning

## KEYWORDS

ethics codes, social movements, digital differential vulnerability, digital vulnerability, data science, public interest technology

## 1 Introduction

Concerns about the power and responsibility of data technology spurred the recent simultaneous publication of multiple ethics codes. Organizations that published data ethics codes recognized the growing need to articulate the technology's potential benefit to society or the "social good". We report on a critical discourse analysis of ethics codes written between 2015-2019 about data science, machine learning, computer science, and artificial intelligence (AI). We use the term data technology to encompass all data-intensive research. We interrogate how ethics codes promote benefits across society. Drawing on a classic provocation [9] for social science researchers to reflect on their position in society, we frame our investigation by asking: whose side are the ethics codes on?

Challenging the assumption that the public good is readily obvious, we argue that all populations are not uniformly considered in ethics codes and, more troubling, ethics codes do little to support vulnerable populations. A vulnerable population is defined here as a group that has been historically and systematically disenfranchised in addition to those currently experiencing a marginalized status, as defined in the Belmont Report on research ethics [46]. We introduce the term "digital differential vulnerability" to explain the spectrum of population experiences with data technologies.

Current disputes about the fairness and accuracy of data technology reveal tensions in defining the benefits across society. Vicious attacks drive female users off platforms designed for free speech [15]. Cambridge Analytica gathered the social networks of casual online labor workers [27, 14] disproportionately targeting people who rely on unstable employment. Facial recognition software in photo applications humiliates users by auto-tagging their photos with animal names [48]. These examples reveal disagreements over the equal distribution of benefits from data technologies throughout society. Many of these examples came to light concurrently to the production and publication of the ethics codes in this study.

Our research design employed multiple methods to focus on language shared across a set of ethics codes. We grounded our methodology in inductive qualitative coding and critical discourse analysis to capture meaning from 15 ethics codes written between 2015-2018. We contrasted these findings to interviews with community organizers who have lived experience of these technologies. A quantitative corpus analysis verified inductive findings and compared the five ethics codes written by corporations to the ten written by professional associations.

The language in this set of corporate and association ethics codes conveyed a sense of duty to society but also narrowly addressed clients and customers. Our results challenge the assumption that ethics codes consider concerns of society at large. The current generation of ethics codes mask differential vulnerability and the contested nature of shared resources, including the social good. We underscore this point around disproportionate exposure to risk and potential harm by interviewing activists and organizers who identify with a historically marginalized community and who rely heavily on these technologies.

We argue that ethics codes that elevate the concerns of customer populations may also subordinate the needs of vulnerable populations creating a two-tiered system of social value. We suggest that ethics codes that want to establish credible moral leadership would extend a sense of commitment and accountability to vulnerable populations.

## 2  Background

Contemporary conversations about algorithmic fairness reveal contested values over social benefits. Are we doing good for everyone, for most people, for the vulnerable, or for a power elite? Critical theorist Sylvia Wynter's "master code" [66] on valuations of human-ness enables us to examine differential vulnerabilities within the current technological landscape. In the case of ethics codes, we question whose interests are represented and which lives are valued and, by corollary, possibly devalued.

### 2.1  Whose side are we on?

The sociologist Howard S. Becker [9] asked the enduring provocation "Whose side are we on?" to encourage researchers to reflect on their position in the society being studied. Becker's piece, a classic expression of the need for reflection, has echoed in the decades since it was written in 1954. The management scholar Donald Schön [57] led the Reflective Practitioner movement in the 1980s. Becker's inquiry also has been essential for contemporary research [59, 60]. We argue that Becker's inquiry is pertinent when using data to understand society.

Becker asserted that established social order makes it impossible for anyone to conduct research on human populations that is completely "uncontaminated by personal and political sympathies" [9]. Instead, Becker asks researchers to equally trace the reasoning of those outside and inside dominant power structures. Those outside may have a justified suspicion of the current system while those inside may have investment in perpetuating it. Reflecting on contested digital resources is especially central to the emerging field of public interest technology.

Becker's hierarchy of credibility [9] frames our argument. The hierarchy of credibility acknowledges a differential distribution of "credibility and the right to be heard" between superordinate and subordinate groups within social systems [9]. At the top of the hierarchy, superordinate groups may be likely to have access to broad information, drive the goals of social systems, and also have the power to implement them. Becker urged researchers to question knowledge claims made by superordinate groups and their right to define what is widely beneficial.

Towards the lower end of the hierarchy, subordinate groups have less strategic information about the social system but have deep expertise through their lived experiences. Their evaluations can be dismissed as complaints and do not become widely established knowledge claims. Without any explicit authority or without organizing into collectives, their perspectives may only be heard in a few isolated voices, making it difficult to challenge institutional systems.

The controversy over risk assessment scores [1, 63] demonstrates how differential positions contest whose beneficence or social good is at stake in data technology. The algorithm COMPAS, Correctional Offender Management Profiling for Alternative Sanctions, assessed the risk of people who had been arrested. Superordinate groups, such as the Wisconsin Department of Corrections, may have purchased the COMPAS system for operational efficiency and increased public safety. Subordinate groups, such as low-risk individuals in neighborhoods with high arrest rates, may be incorrectly assessed by COMPAS and incarcerated to protect public safety. COMPAS is an example of how power, the capacity to value and devalue lives, operates when superordinate and subordinate groups disagree over shared resources. Becker instructs researchers to avoid bias by tracking the reasoning of all groups.

Ethics codes for data technology are often explicitly concerned with preventing bias, for instance the ACM 2018 code contains language to "be fair" [6]. Batya Friedman and Helen Nissenbaum [22] refer to pre-existing, technical, and emergent bias in computer systems as systematic and unfair discrimination. In this study we investigate whether the ethics codes engage with Becker's hierarchy of credibility.

## 2.2 Contested Values

We connect Becker to Sylvia Wynter's conception of the "master code" [66] which functions as a means of organizing society into a hierarchal order. Wynter helps us consider who is included in ethical codes' consideration of society to better understand the question of 'who' benefits from social good and 'who' bears the risk of harm. Grace Kyungwon Hong [31] extends Wynter's 'master code' [66] considering how it legitimizes death of some to maximize the lives of others. To put in another way, categories of protected life imply a category of unprotected life.

Lawrence Lessig [39] further reminds us to interrogate power and social order in technological architectures given that the process of technological design is also a political one. Lessig suggests that politics is a process of reasoning about ideals. Ethics codes seeking to define ideals for society naturally insert themselves into politics. Current debates around the harms and benefits of data technology reflect a political process that is reasoning about ideals.

Differential experiences of harm on social media platforms provide some additional examples. Users have challenged platforms over the contradiction of their free speech advocacy and their policies governing harmful speech. In an open letter to Facebook, Data for Black Lives founder Yeshimabeit Milner [44] requests that Facebook establish a Data Code of Ethics by arguing that structural inequalities can be codified in new data-driven technologies. Existing terms of service do not adequately protect all users given differential exposure to online harassment and violence based on race [54, 55], gender [15, 61] or other categories. People with these experiences have limited options for protection within platform policies [25] and recourse through judicial systems [5] leading to a legitimate perception that their communication and commerce are less valued than other users [21]. Because technology corporations are ubiquitous without alternative competition [28] in their control of information infrastructure, vulnerable populations rely on these technologies for day-to-day practices as heavily as others, but unevenly experience risk and harm.

Defining what is good for society and the wider public can create new vulnerabilities and perpetuate old ones. Importantly these dynamics are often experienced at the population level and not the individual level.

## 2.3 Codes of Ethics

Ethics codes are documents that function as an important symbolic commitment to society and reflect a set of moral principles and values. Occupations like engineers and doctors follow a professional code of ethics [46] that establishes general guidelines that outline their moral obligations to building structures and treating patients. Critical data studies [47] and data scientists [52] equally are seeking how to develop similar ethical structures for this new field.

Early efforts to produce technology ethics codes emerged in the 1970s along with the movement that equated software with engineering [3]. The recognition of software's legal liability and growth of corporate social responsibility led to the development of additional ethics codes in the 1990s. This was an important recognition of technology's leading role in business, and wider adoption in consumer households. By 2010, rising concerns about big data and privacy lead to more calls for ethics in the industry. Ad-hoc meetings of academics and leaders published documents on ethics [23] while prominent professional associations revised existing ethics codes to keep up with new technologies [6]. By the time this conference on Fairness, Accountability, and Transparency was founded in 2014, there were regular meetings [65] to address public concerns about data technology. Our research draws on documents produced from 2016-2019 efforts.

An uncontested public benefit is a convenient assumption underlying most ethical codes. We want buildings that do not fall down, bridges that can withstand hurricanes, and doctors who do no harm. The history of social science [41] exhibits many well-intentioned projects that answer valuable scientific questions to serve society yet harm participants. Historically, many of these participants experiencing harm were not powerful groups in society [20]. It becomes clear from these examples that the good of the society at large is often in opposition to the lived experience of specific individuals or populations.

Research ethics emerged as a solution to state-sanctioned experiments on vulnerable populations in Europe [58] and the United States [20] during the twentieth century. However not all ethics codes are straightforward. Most governments publish an ethics code for research involving human subjects [56] yet also review every medical or behavioral experiment.

Our study challenges the underlying assumption of a single public benefit within ethics codes published by corporations and associations. Professional associations, such as the ACM, Association for Computer Machinery and the IEEE, Institute of Electrical and Electronics Engineers, regularly produce codes of ethics [5] to mitigate occupational risk. Individual organizations issued their own code of ethics to publicize their self-governance efforts [26, 45]. We note that several popular technology companies, such as Apple, Amazon, Facebook and Twitter, have not yet issued a code of ethics [21, 27]. Comparing corporate and professional associations allows us to see how different organizational authors conceptualize the public good.

## 3 Methods

This project investigated issues of power, responsibility, and the social good by asking: What is the aspirational social obligation that ethics codes encourage? We follow a logic of inquiry where language is constitutive in shaping knowledge, culture, politics, and other social orders. Qualitative textual data gathered in a natural setting are ideal to learn about emergent meanings [28, 40]. Interviews of community organizers and activists at the margins of the technology infrastructure supplemented the discourse analysis of ethics codes. We used both positive and negative evidence to establish our findings. Our methods connect the dots between distant verbal ethical framing in the documents and the intimate oral ethical framing as experienced by marginalized communities. Combining the ethics code analysis with the unique perspectives of organizers and activists led to our discussion of digital differential vulnerability.

### 3.1 Research Design

Our approach follows an interpretive qualitative research methodology that employs inductive logic to analyze textual material written in documents and generated through interviews. Discourse analysis alerts the researcher to patterns of language associated with different social purposes such as outcomes, tasks, events, or populations. Critical discourse analysis [18] focuses on understanding how politics and power dynamics are represented in language. Qualitative researchers were attuned to language that connected specific actors (such as data scientists, designers, clients, stakeholders, etc.) to verbs that denote actions. Focusing on actors and actions directed our consideration of who holds power and responsibility in determining social good and how costs and benefits are distributed across different stakeholders.

After a small team conducted a preliminary reading of the ethics code in summer 2018, we assembled a team of seven qualitative researchers during summer 2019. Each document was coded independently by each member of the team. Before conducting a thematic analysis, each coder familiarized herself with the structure, flow and topic of the each document. We used Dedoose Version 8.0.35 to manage and organize the code book and to share our coding. Through weekly coding discussions, we considered the differences in language and type of organizational author. We incrementally moved from open coding to axial coding [40] in an iterative coding process.

Our close reading was bolstered by a quantitative corpus analysis of the 43,771 words in the ethics codes set which provided a critical distance to our analysis. Distance is an essential part of establishing validity in qualitative studies [53]. We investigated concepts further through collocation and word frequency analysis. All frequency counts excluded a set of limited stop words. Whole words and partial words were identified using regular expressions, and word sets were based hypernyms from WordNet 3.1. With both qualitative and quantitative approaches, we are better able to confirm intuitions about language patterns.

In conjunction with a long-term research project, one author conducted 60-minute interviews with 30 community organizers and activists with expertise in technology development, network coordination, and digital media strategy. Working on campaigns such as policing, mass incarceration, and immigration, interview participant discussed the relationship between technology and social change. As social movement organizing relies upon the market logics of the digital information and knowledge economy, the effects of autonomous systems in information distribution impact the efficacy of campaigns for social change. Participants are identified with a single letter along with the date of the interview. We completed Institutional Review Board certification for the interviews but the public ethics documents did not qualify for additional review. We include reflections from these interviews that resonate with topics of power and social good.

### 3.2 Professional and Corporate Ethics Codes

We selected five corporate ethics codes and ten codes written by professional associations. Ethics codes for this study are defined as short documents with a statement of principles usually accompanied by additional text. To be included for analysis, a group or organization must be the author and the topics were limited to machine learning, computer science, data science, AI or analytics. As part of our selection criteria, documents also needed to enumerate specific principles. We eliminated documents with stated principles by individual authors that did not represent an organization, as well as statements of principles published as journal articles [67, 68]. Principles framed as organizational goals, such as the Partnership for AI's thematic pillars and tenants, were also not considered. The selected codes by corporate authors roughly matched the size of the codes by associations.

We identified ethics codes for analysis through in-depth literature analysis and search. In addition, we reviewed aggregate efforts such as Doteveryone's crowd-sourced list of oaths, pledges, and manifestos [36]. Our initial list was generated out of necessity as one author was repeatedly requested to attend convocations to write ethical codes and began to keep a list of documents out of curiosity about the number of efforts. This list was regularly augmented.

Association ethics statements analyzed in this study include documents authored by: ACM [3,4,5,6] Algorithmic Justice League [7], Association of Internet Research [00], Future of Life Institute [23], Data Science Association [17], Institute for Ethical AI & Machine Learning [34], IEEE [32], and the Japanese Society for Artificial Intelligence [30]. Corporate documents include Google [26], Intel [35], IBM [33], Microsoft [45], and Axciom [8]. See Appendix for a complete description of codes reviewed in this study.

## 4 Findings and Analysis

Our inductive research sought to identify how ethics codes defined social good and how they addressed power and responsibility of technologists. We looked for similarities across all ethics codes while also seeking differences between the corporate and professional association codes.

The 15 ethics codes contained 43,771 words total, with an average of 2,918 words per document. The shortest code of ethics is the Asilomar AI Principles [23] since it contained only a list of principles with no additional statements. The word count distribution is nearly equally divided, with associations making up 55% of the corpus and corporations 45%. This means that the five corporate documents are nearly the same word count as the ten from associations. The five corporate codes total 20,250 words and average 4,050 words each, and the ten professional association codes total 23,521 words and average 2,352 words. Using an n-gram analysis, we discovered that the only common noun phrase amongst all 15 ethics codes was "personal data."

Three core themes that emerged from the interviews and documents were: duty and responsibility, harm and bias, and social good. Below we review how these themes develop and draw on Becker's model of superordinate and subordinate groups [9] to approach dynamics of power in our evidence. In this case, we locate government organizations, professional associations, and corporate entities as superordinate groups, who hold power and responsibility in controlling systems by defining constitutes good and what constitutes harm. Subordinate groups, such as users and community organizers, experience the systems established by superordinate groups.

## 4.1    Duty and Responsibility

All codes had a sense of duty and responsibility in common. Association codes reflected duty to the profession, which includes an obligation towards the public interest that promotes occupational integrity. The most common words in the professional association documents were: data, should, research, ethical, scientist, information, systems, client, computing, and human. This reflects an emphasis on the scope of work of technologists and computing occupations. Because associations aim to protect professional reputation, their ethics codes were more careful to delineate responsibility and accountability in the case of harm. As a whole, ethics codes elevated the agency of technical systems and individuals with minimal consideration of the role of organizations who employ technologists.

In the corporate documents, the most common words were: data, ai, will, should, we, they, bot, privacy, information, consumers, ethical, and you. This reflected a product-forward focus, emphasizing the technology itself and how it impacts society, as well as indicated a consumer relationship between company and user. Corporate codes tended to position consumers as a subordinate group lacking knowledge of data collection processes and algorithmic decision-making while positioning themselves as trustworthy stewards on the path towards developing social good [8].

The qualitative analysis brought up the recurring use of consumer, customer, and client. We explored this cluster of words in the quantitative analysis. The consumer cluster appeared 122 times across the five corporate ethics codes most frequently collocated with: existing, view, tools, help, and data. This indicates a connection between customer use of technologies and social benefits. By contrast, the consumer cluster appeared only 27 times across the ten association codes. The associations collocated the terms with stakeholders they feel a sense of obligation and accountability towards: affiliates, employees, parties, and client. The words were not equally distributed across all documents. The Data Science Association used the term "client" the most [17], and Axciom used the term "consumer" the most [8].

In professional codes of ethics, words associated with terms about people appeared 1791 times across 54 unique words. The words most associated include: dignity, private, subjects, rights, good, relevant. Since documents generally focused on professional responsibility, this indicated a sense of duty and obligation to serve people's well-being.

In corporate codes of ethics, words associated with terms about people appeared 1560 times across 48 unique words. Words most associated include: policy, sensitive, bots, access, information, new, and if. These words indicate considerations for how technologies may impact people and how to govern new technologies.

## 4.2    Power to Harm and Benefit

Despite a theme of duty and responsibility, the ethics codes tended to be vague around practices, processes, and agents. Google references an amorphous 'we' that operates in the future tense: "we will design", "we will test", "we will work", "we will responsibly share". Intel refers to AI technology as an independent entity [26]. Anthropomorphizing technologies, it can be argued, creates ambiguity around accountability, control, and power [6, 45]. While corporations and associations tend to claim responsibility towards bringing social good, their ethics codes reflect muddiness around who (or what) should be held accountable.

Many of the ethics codes include discussions of conflict and tension, such as "harmful bias in the design"; "errors in data bias"; "implications for physical safety"; and "harmful discrimination." As ethics codes take on a technocratic position, these conflicts have not been framed as openly political. Ethical documents presume an apolitical hierarchy between technology organizations in relation to users, professionals and clients, and policy-makers and citizens. By rendering harm and bias as apolitical, rather than connected to pre-existing social and political problems, harmful impacts of technology become something that superordinate groups can "monitor", "avoid", and "prevent" as written in the ethics codes. The codes assume that a neutral position of both organization and technology can be possible.

In addition to the lack of precision regarding agents responsible for particular actions, the specific uses of technologies in producing social harm and good also remains unclear. The vagueness of the language makes the practical uses of ethics codes difficult. The ethics codes did not provide clarity around harm, benefit, and agency to act.

## 4.3  Differentiated Good

Society is a broad concept that many groups lay claim to. The ethics codes aspire to have a positive impact on human lives. Community leaders, organizers, and activists aspire to social change using digital technologies in campaigns and movements to shift structural and material systems. Interviews about social change supplemented the ethics code analysis to better understand how social good is differentiated. These interviews are part of a long-term ethnography on social movements and digital technologies. The cases, stories, and observations reflect the experiences of organizers who work with and live in marginalized communities. We report on the initial findings of these interviews to bring into focus the gulf between the ethics codes and lived experiences.

Organizers and activists can be considered as users and therefore part of the constituency identified in corporate ethics codes. As users, they should be served by corporate duty and obligation towards a common good. However, they tend to be excluded from superordinate decision-making processes around technological governance and development. They represent the interests of subordinate communities and people most impacted by inequalities in data surveillance and algorithmic decision-making, such as the uses of algorithms in sentencing guidelines. Working closely with community members, they are accountable to the communities they represent and often, they are part of those communities themselves.

Community organizers raised concerns around data surveillance, information bias, consent, and safety in conversations about design values [22] through constraints and affordances of technologies. Because the interviewees primarily employed corporate technologies for community organizing due to reasons such as access and resources, our discussion emphasizes the shortcomings of corporate platform technologies as a predominant example. A recurring theme among interviewees highlighted that users have both differential experiences of social good as well as different levels of vulnerability to harm and bias—that those most at risk for harm also do not fully experience benefits.

Three key points from the these interviews demonstrate differential experiences: 1) despite prioritizing consumers, attempts at technological neutrality fail to address disparities in costs and benefits across user groups; 2) potential profit overrides other principles and values; and 3) technological solutions are rarely sufficient to create social change. Together these differential experiences of social good and personal harm dispute the assumption of a unitary public benefit.

*4.3.1 What is Neutrality?* The myth of neutrality in technological platforms obscures how pre-existing biases in the surrounding social world can be replicated in technological practice. As Becker and Wynter point out, society is already organized into hierarchal orders where some lives are more protected than others.

The first example highlights contested values alongside differential experiences of vulnerability. B. works as a network coordinator and organizes against caste oppression and religious intolerance. Twitter CEO and co-founder Jack Dorsey met with the Dalit community about their safety as Twitter users. B. describes the aftermath of a photo of that meeting.

> "The rise of the Hindu fundamentalist trolls in South Asia makes it really hard to talk about caste on Twitter. When Twitter CEO Jack Dorsey held up a poster that said 'Smash Brahminical Patriarchy' that [we] designed, Hindutva trolls attacked Twitter and Twitter's lawyer, also an Indian upper caste person, both issued apologies and took down the photo right away. Mind you, they were the ones who took the photo." (B., March 2019, Interview #14)

Twitter's legal head, Vijaya Gadde, issued an apology to those offended by the photo and stated that Twitter failed to live up to its goal of being an impartial platform for all users. Organizers from the event stated that the platform had a responsibility not only to promote free speech but to guarantee safety for most the vulnerable users.

> "[We] sat with Twitter to talk about it as an unsafe space for Dalit activists, for people who are bringing up caste apartheid, and yet in the end, Twitter took the side of upper caste trolls...The caste system is 3000 years old. We've had these hierarchies and they're being digitized." (B., March 2019, Interview #14)

B. perceives Twitter's statement of impartiality as one that reinvests in existing hierarchal systems. In the society where B. lives, there are clear differences in social and political power between castes that also create differences in experiences online. B. and other activists suggest that in order for corporate platforms to act responsibly towards all users, they must actively pursue practices to protect users that have been historically marginalized in society.

Another organizer, C., discusses how the lack of online safety can translate into other forms of harm. She shares an experience where her participation in a Twitter forum on sexism and misogyny within Asian American communities made her a target of harassment:

> "After the conversation, I was on some reddit threads— someone shared my info...my telephone, email address, where I worked...It's scary because anyone can walk in and find me. Online toxicity doesn't just exist online it seeps into physical spaces, where you work and live. All of that is really frightening."  (C., November 2018, Interview #3)

While corporate ethics codes point to "risk of harm" or "potential for harm", there is little consideration of how users may be disproportionately exposed to harmful consequences.

Both these stories demonstrate that users become exposed to harm unevenly. Their stories also reveal how assumptions of technological neutrality and user equality can harm those without existing power within society. Codes of ethics must attend to the ways that society is already structured and the experiences of disenfranchisement within those structures.

*4.3.2 Whose Values?* Across interviews, organizers felt that corporate platforms tended to prioritize potential profit over humans and human experiences. One organizer pointed out the ironies of how Facebook derives profits from events that generate viral circulation of violent images. While corporate ethics codes may emphasize moral authority and technical expertise as part of branding, organizers expressed concern that ethical principles become undermined by a corporate bottom-line.

> "We in some way have placed this expectation of morals on these platforms that are based in capitalism…These tools are tangled around expecting big companies to have progressive views…however, [corporations] see all harm as the same extent of harm." (F., December 2018, Interview #11)

> "There needs to be a radical shift in our value systems and not putting profit over people…I've been thinking about ways that we can create a platform that isn't driven by capital and investments and by the amount of money we can potentially make." (C., November 2018, Interview #3)

Both F. and C. emphasize that while corporations may promote a desire to serve the common good, there remains a tension between profit and the public good. Ethical statements like 'do no harm' and 'serve the public interest' become tricky to interpret when linked to for-profit operations. Another organizer shared concern over private-public partnerships, such as Amazon's cloud services which enables data sharing between law enforcement and social media platforms. One organizer noted that social change events publicized as Facebook events are not widely promoted on newsfeeds unless an ad is purchased. Requiring payment to digitally distribute information advances the knowledge claims of those, often from superordinate groups [9], with the capacity to pay. This conversely unintentionally suppresses events sponsored by those with fewer resources.

Organizers' wariness towards corporations' creates a sense of distrust and also frustration that they need to rely on corporate tools in their pursuit of social change. They also articulated an awareness of their position as users of these platforms.

> "I really dislike and distrust organizations like Google, Facebook, etc. and feeling like I don't have alternative but have to use it anyway." (J., March 2019, Interview #17)

> "We're ultimately customers of these platforms…we're both organizer and customer." (P., June 2018, Interview #4)

Often users are narrowly conceived as a consumer with purchasing power. While publicly traded companies have the right to focus on profits and build shareholder value, prioritizing consumers over others in society legitimizes a hierarchy of who is valued and devalued.

*4.3.3 Other Solutions?* Community organizers suggest that part of platforms' lack of accountability to vulnerable user groups stems from an absence of deep relationship building with these groups. While corporations may take into account external perspectives, such as Dorsey's meeting with caste abolitionists, organizers feel that there are insufficient grounds to build a relationship of trust and accountability. V., a designer who works on building more consensual technologies, says:

> "The struggles are not just an issue or intellectual problem, but you see how people are impacted by those struggles on a day-to-day basis." (V., March 2019, Interview #13)

Rather than being stewards for users, technologists building solutions to societal problems should develop more accountable relationships with directly impacted communities. In approaching social good, corporate ethics codes approach "challenges to society" as something to be solved through technological empowerment. Rather than imagining social good as a "cool creative problem to solve", V. argues that designing for more just societies requires more relationship building processes than technological innovation.

Google's 2018 statement emphasizes that the future potential of technologies to "improve our lives is profound." But, the question remains, whose lives? Several organizers shared skepticism that technological innovation for good would fix existing social problems. Rather, they felt that technological solutions would continue to perpetuate social stratifications.

> "Why is it that people who haven't had access still don't?" (C., November 2018, Interview #3)

> "People who are most impacted [by climate change] are poor Black and Brown people. Already tech companies like Google and Facebook are working with the fossil fuel industry to come up with false solutions like contraptions that can suck carbon out of air. What will happen is most elites will be fine…those who can't access it will be most impacted by climate crisis." (I., November 2018, Interview #6)

Findings from these interviews illustrate that people experience technological costs and benefits unevenly based on existing social hierarchies. Through the example of Twitter's responses to community safety; critiques of corporate value systems; and observations on the limits of innovation, these interviews with organizers challenge understandings of power and duty in traditional technology development.

In connection with findings from the text analysis, organizers demonstrate the limitations of data ethics codes. The evidence in the interviews questions framing users as consumers and calls attention to the ambiguity of accountability when harm occurs. These interviews reflect the varying scales and registers of experiences of users that are certainly not imagined within corporate ethics codes.

## 5   Discussion

Insert text here for the enunciation or Math statement. Insert text here for the enunciation or Math statement. Insert text here for the enunciation or Math statement. Insert text here for the enunciation or

### 5.1   Customers are not society

The consumer-orientation in data ethics documents reflects a concern for societal well-being that slips into direct obligations towards customer bases. While the codes intend to address social concerns broadly, they use language that targets only their users. The opposite of users based on our evidence is not non-use of the technology but a duty towards the general good and society as a whole. While corporations serve populations that rival the size of small nations, they still cannot adequately serve broader society if they only value their potential and current users. We argue that credible moral leadership extends beyond a population identified with profit and should serve society broadly.

People using technology as a means to achieve social change belong to a specific subset of users served by the technology industry and professionals. Activists and organizers use digital technologies to achieve their vision, such as relying on autonomous information systems to distribute campaign messaging, using messaging platforms to mobilize political actors, or questioning whether the data industry perpetuates societal harm. Yet, as our interviews show, activists-as-users do not feel as if platforms truly serve them or the communities they represent. When users are only valued based on their consumer potential, this excludes the interests, needs, and concerns of people in society using those systems to achieve social, political, or non-commercial goals.

The organizations and associations writing ethics codes might consider whether they are serving only superordinate voices who might dominate discourses over what is good and what is harmful. This might be particularly important when considering who deserves [5] to experience costs and benefits. An awareness of subordinate perspectives might expand ethics codes towards definitions of benefit and harm that support more people.

### 5.2   Digital Differential Vulnerability

The assumption that there is an obvious single idea of what is good for society oversimplifies the task of solving issues of fairness, accuracy and transparency. This assumption, as shown in our evidence, often avoids marginalized groups that have an equal claim to resources and protection. Revelations of bias in machine learning [11, 12] are evidence of contestations over digital resources. We extend Wynter's [66] argument that categories of protectable life in social and economic systems imply categories of unprotectable life to technological systems. Our observations suggest that technological systems reveal differential vulnerability through digital representations.

We introduce the concept digital differential vulnerability to describe how vulnerable populations are disproportionately exposed to harm through data technology that seeks to promote a single point of social good.

Communication scholars use digital vulnerability to refer to behavior that makes someone susceptible to harm. Vargas [62] defines digital vulnerability within the context of exposure to criminal justice technology, referring to citizens' risk of having incriminating information publicly disclosed and exploited by third parties. We broaden that definition to consider that some populations have less ability to avoid these systems [11] and, in addition, are compromised [12] by being included.

The two sets of evidence in this study both claim to support society, whether through creating professional standards or through social movement organizing. Unimagined users encompass a wide range of different people across geographies, experiences, communities, that are often unequal and incommensurable. If ethics codes that merely present moral aspiration fall short, the moral goalposts for the actual data-intensive work itself may be compromised. An awareness of digital differential vulnerability can push technology developers to reshape their assumptions about the user base and critically engage with the complex dynamics of harm and good across populations.

### 5.3   Recommendations

Ethics codes are not static and efforts to revise and improve the current set of documents are likely to continue. We provide three recommendations that consider digital differential vulnerability in the development of data ethics codes.

**1. Differentiate between business values and a mission to society.** The normative mission of most ethics codes is to meet an obligation to society. Discussing protection of vulnerable populations may be a way to demonstrate support for the public good. Organizations writing these codes may consider creating separate documents specifically for their client base.

**2. Clarify power and agency.** We noted that there was little discussion of who actually holds power over data-driven technology. While obvious in the case of most professions (i.e the doctor has control over the patient), agency is less clear in this occupation. Some advice in the ethics codes was clearly addressed to a consultant who has the ability to refuse yet many technologists work within an organization that has ultimate control over their actions. Professional associations might begin to establish a position on how an orientation to the organization may impact a technologist's level of ethical agency.

**3. Consider whether a societal goal is attached to an existing political process.** Many data technologies cross geopolitical borders and legal jurisdictions yet address issues that are usually negotiated through a political process. Ethics codes that attempt to be universal must recognize cultural variations to issues like free speech, housing, or healthcare. Populations experiencing digital vulnerability may not have a voice in normative political processes and may experience additional harmful experiences. In those cases, technologists might consider working within and alongside specific communities to see if a broader political solution might be necessary before a technology solution could take hold.

An ethics code that meets these goals is the Algorithmic Justice League's Safe Face Pledge, which is document that asks organizations to make a public commitment about the ethics of facial analytics [7]. The Safe Face Pledge recognizes specific risks before assuming beneficence for all and situates agency within organizations.

Public interest technologists will be seeking to successfully act on their concerns for the greater society. The ethics codes overall failed to provide substantive advice on what to do or what types of situations this codes even apply in. Some efforts have promoted a checklist approach [42] to support better practical advice. Technologists within and outside organizations could use clearer and more specific guidance on how to handle difficult situations. This need is evidenced by efforts to institute a code of ethics initiated by working professionals, such as the 2018 crowd sourced Community Principles on Ethical Data Sharing (CPEDS).

Ethics codes sparked a vigorous conversation around data ethics yet these efforts take only the first steps towards building a professional consensus. Data technology is constituted and constitutive of a range of power relations from individual workers to government regulators. The production and consumption of technologies extends well beyond ethical principles. The datasheets for datasets [24] solution, which argues for additional documentation to accompanying digital material, echoes Becker's call for researchers to be transparent about themselves and their data. More research on public interest technology may begin to identify ways to extend benefits of safety and security more widely.

Future studies may interview technologists to understand what they need in order to concretely pursue ethical work in practice. Other potential directions may include careful analyses of documents written by activists and organizers, such as Data for Black Lives' letter to Facebook [44] or alternative policies on harmful speech published by the Change the Term Coalition (www.changetheterms.org). Public interest technologists might work with these organizations to develop the next generation of ethics codes. Additionally, while we included several multinational firms and different national professional associations, we did not truly address the global nature of the technology industry. Further studies might take a more transnational approach to compare sociocultural and legislative contexts across geographic regions, such as how notions of data privacy are considered between the African Union, the Association of Southeast Asian Nations, or the European Union.

## 6   Conclusion

Our study challenges the idea of a single public benefit by pointing to hierarchies embedded in technologies. The data ethics codes in this study strived to address the concept of social good through a sense of duty but primarily focused on the needs of visible users. Interviews with community organizers seeking social change surfaced the limits of technical solutions to concerns of marginalized communities. Together, this evidence highlights the distance between the ethics code documents and lived experiences. Digital differential vulnerability explains disproportionate exposures to harm experienced through data technology. We argue that ethics codes that address social concerns about technology should consider the specific needs of multiple communities in order to more broadly reduce harm and more evenly distribute benefits.

## ACKNOWLEDGMENTS

This project would not have been possible without the tireless efforts of the 2019 QLab. The outstanding research assistants on this project, in alphabetical order, were Shannon Kay, Chloe Marten, and Molly Nystrom. Dr. David C. Morar and Jessica Spencer provided technical assistance. Kiran Samuel, Shawn Janzen, and Kandrea Wade made additional contributions. Thanks to the anonymous reviewers. Professor Washington also extends gratitude to all NYU students in the Spring 2019 Ethics of Data Science course.